\documentclass[runningheads]{llncs}
\usepackage{graphicx}
\usepackage{pgfplots}
\usepackage{float}
\usepackage[inline]{enumitem}
\usepackage[nolist,nohyperlinks]{acronym}
\usepackage[]{todonotes} %
\usepackage{csquotes}
\usepackage[style=lncs,giveninits=true,maxnames=3,mincitenames=1,maxcitenames=1,backend=biber]{biblatex}
\usepackage{tabulary,lipsum}
\usepackage{microtype}
\usepackage{booktabs}

\usepackage{xurl}
\usepackage[hidelinks,bookmarks=false,final,breaklinks]{hyperref}
\usepackage{listings}
\usepackage[lighttt]{lmodern}
\usepackage{tabularx}

\newcommand{\repeatthanks}{\textsuperscript{\thefootnote}}
\usepackage[absolute]{textpos}

\begin{document}
\begin{textblock}{14}(1,0.1)
\begin{center}
\noindent N. Boltz, S. Hahner, et al., "An Extensible Framework for Architecture-Based Data Flow Analysis for\\
Information Security," in European Conference on Software Architecture (ECSA), Springer, 2024. 
\end{center}
\end{textblock}

\title{An Extensible Framework for Architecture-Based Data Flow Analysis for Information Security}

\titlerunning{An Extensible Framework for Architecture-Based Data Flow Analysis}

\author{Nicolas Boltz\thanks{Both main authors contributed equally.} \and
Sebastian Hahner\repeatthanks \and \\
Christopher Gerking \and
Robert Heinrich
}
\authorrunning{N. Boltz, S. Hahner, et al.}

\institute{Karlsruhe Institute for Technology (KIT)
\email{\{boltz,hahner,gerking,heinrich\}@kit.edu}}

\maketitle

\begin{abstract}
The growing interconnection between software systems increases the need for security already at design time. 
Security-related properties like confidentiality are often analyzed based on data flow diagrams (DFDs).
However, manually analyzing DFDs of large software systems is bothersome and error-prone, and adjusting an already deployed software is costly.
Additionally, closed analysis ecosystems limit the reuse of modeled information and impede comprehensive statements about a system's security.
In this paper, we present an open and extensible framework for data flow analysis. 
The central element of our framework is our new implementation of a well-validated data-flow-based analysis approach. 
The framework is compatible with DFDs and can also extract data flows from the Palladio architectural description language.
We showcase the extensibility with multiple model and analysis extensions.
Our evaluation indicates that we can analyze similar scenarios while achieving higher scalability compared to previous implementations.

\end{abstract}

\section{Introduction}\label{sec:intro}

As our modern world becomes increasingly digitized, the integration of various digital services into our daily lives has become more prevalent. 
To enhance the quality of service, a growing amount of data is stored and processed, e.g., online shops utilizing purchase history data for recommendations. 
The seamless exchange of such collected data between different services or systems is a common practice. 
In scenarios like online shopping, sensitive information like payment details and customer addresses are involved. 
Consequently, security becomes a central concern in designing and building such software-intensive systems.
 
Information security has several definitions, e.g., as the CIA triad of confidentiality, integrity, and availability, or in ISO 27000 \cite{isoConfidentiality}.
More recent legal regulations, like the General Data Protection Regulation (GDPR) \cite{GDPR}, define information security more broadly.
For modern systems, changes and reconfiguration in the context, environment, or internal structure might occur frequently \cite{weyns23researchagenda}.
Since the protection goals are highly dependent on the system under consideration, the protection goals that must be addressed may also change. 
In addition to the CIA goals, other protection goals might be considered, like privacy, authenticity, non-repudiation, accountability, and auditability. 
A system violating confidentiality or privacy can cause costly fines, as seen in the case of H\&M~\cite{the_hamburg_commissioner_for_353_2020} or British Airways~\cite{british_fine}. 
However, identifying such violations can be difficult, because the interconnected software systems represent complex networks of data flows. 
Hence, a holistic and scalable approach to analyzing them is required.


Data flow analyses based on source code, e.g., JOANA \cite{Joana}, KeY \cite{key}, or CodeQL \cite{codeql}, cannot consider context information, such as deployment. 
However, such information can be essential for information security, e.g., whether the application is deployed to an external cloud provider or not. 
In addition, source code analyses cannot be used in early design phases because of their need for existing source code. 
Analyzing the system during design time is beneficial because fixing issues in later phases is usually more costly \cite{shull_what_2002}. 
Seifermann et al. \cite{PalladioIntegrationConfidentiality, seifermann2022detecting} proposed an architecture-based data flow analysis to analyze software systems for confidentiality violations. 
Their approach considers additional context information, such as the deployment, enabling software architects to analyze confidentiality during early design phases. 
However, the original Prolog-based implementation of \textcite{seifermann2022detecting} is hard to maintain and has a high resource demand, which severely limits the applicability for large software systems.
Although they already used a model of a data flow diagram (DFD) \cite{demarco1979structure} as an intermediate representation during their analysis, they did not continue to follow the idea of using DFDs as the primary model artifact. 
With appropriate tool support, DFDs represent a powerful and commonly used mechanism for threat analysis \cite{bernsmed2022adopting} that helps in correctly identifying security-related issues \cite{Schneider24_DFDs_empirical_experiment}.

In this paper, we present an extensible analysis framework centered around our previously presented new implementation of the aforementioned approach to data flow analysis \cite{schwickerath2023toolsupported}. Our framework addresses shortcomings regarding the limited input capabilities, the limited intermediate use of DFDs, and problems with maintainability, scalability, and extensibility:

\begin{enumerate}[label=\textbf{C\arabic*}]
    \item We propose a novel DFD metamodel. 
    In contrast to the previous DFD model \cite{seifermann2022detecting} we do not consider DFDs as intermediate system representations but as primary software architecture modeling artifacts. 
    As part of our framework, we provide means to manually define DFDs as well as automatically derive them from the architecture description language Palladio Component Model (PCM) \cite{reussner2016a} and other third-party diagram representations \cite{Schneider_Scandariato_2023,schneider2023microsecend}. 
    \item We present a new Java-based implementation of the analysis approach of \textcite{seifermann2022detecting}, which is based on a newly developed internal data structure and alleviates the need to mix Java with other technologies like Prolog. 
    In addition, we provide new forms of input, e.g., DFDs defined with our metamodel (\textbf{C1}), and a domain-specific language (DSL) that enables software architects to define constraints or queries for the analysis. 
    We demonstrate how our analysis framework can be applied and extended for other security concerns, e.g., regarding the GDPR \cite{boltz2022model}, or uncertainty \cite{hahnerSeams}.
\end{enumerate}

\noindent
This paper is structured as follows:
\autoref{sec:overview} introduces our data flow analysis framework.
\autoref{sec:dfd} describes our new DFD metamodel (\textbf{C1}). 
In \autoref{sec:analysis}, we describe the analysis (\textbf{C2}), and in \autoref{sec:extension}, we showcase existing extensions.
\autoref{sec:eval} presents our evaluation and \autoref{sec:conclusion} concludes the paper. 
\section{Overview of the Data Flow Analysis Framework}\label{sec:overview}
\begin{figure}
    \centering
    \setlength{\abovecaptionskip}{3pt}
    \setlength{\belowcaptionskip}{-3pt}
    \includegraphics[width=\textwidth]{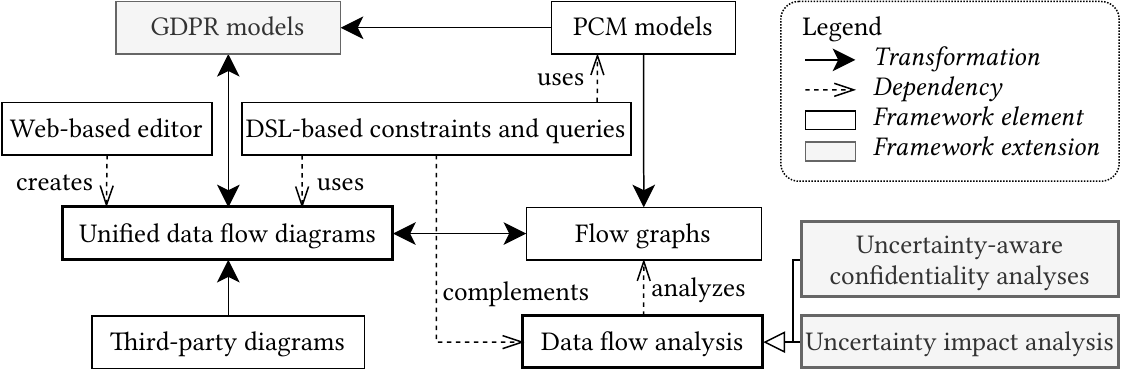}
    \caption{Informal overview of the structure of the data flow analysis framework.}
    \label{fig:overview}
\end{figure}

In this section, we summarize our framework which is explained in more detail hereafter.
\autoref{fig:overview} gives an informal overview of the structure and the dependencies between the different parts of our data flow analysis framework.
We highlight analyses and editors with a bold border; all other rectangles represent models.

\noindent
Unified data flow diagrams (DFDs) \cite{PalladioIntegrationConfidentiality} play a central role in our framework (\textbf{C1}).
These diagrams can be manually defined using a web-based editor or transformed from third-party diagram representations.
Additionally, we enable the extraction of data flows from software architecture models described using PCM \cite{reussner2016a}.
To unify the analysis process (\textbf{C2}), we transform DFD or PCM models into a set of Directed Acyclic Graphs (DAGs), called \emph{transpose flow graphs (TFGs)}.
Each vertex represents one individual data processing operation from either a DFD or PCM model, which simplifies the analysis.
Using label propagation on TFGs, the analysis finds violations of predefined constraints that reflect information security objectives, e.g., confidentiality requirements. 
Constraints and queries can be specified using a domain-specific language (DSL), referencing PCM and DFD models. 
Framework extensions are possible by transforming into our DFDs or TFGs or by inheriting from the data flow analysis.
GDPR models \cite{boltz2022model} are an example of the former while uncertainty impact analysis \cite{hahnerSeams} and uncertainty-aware confidentiality analyses \cite{boltz2022handling, Hahner2023Uncertainty, walter2021architectural} showcase the latter.
The framework is tool-supported and available as open source\footnote{See \url{https://dataflowanalysis.org} or \url{https://github.com/DataFlowAnalysis}}.
This includes modeling support, automated model transformations, and a DSL-supported analysis\footnote{Video demonstration available \cite{schwickerath2023toolsupported}: \url{https://youtube.com/watch?v=q3WJsMyqJcA}}. We also provide a dataset \cite{dataset} including all tooling, code artifacts, and evaluation data.


\section{Modeling and Deriving Data Flow Diagrams}
\label{sec:dfd}
While \textcite{PalladioIntegrationConfidentiality} created a unified DFD notation, they only used it as an intermediary representation for their data flow analysis \cite{seifermann2022detecting}. 
However, as DFDs are an established software architecture representation \cite{bernsmed2022adopting} and are widely used to analyze various types of data security \cite{alshareef22purpose, seifermann2022detecting, sion2018solution, Tuma2019}, we present an explicit DFD metamodel that can also be used as input for our analysis.
The study of \textcite{bernsmed2022adopting} concludes, that, while DFDs are good for evaluating security, there exist challenges in preparing DFDs. 
Especially tooling that improves the effort of creating and maintaining DFDs is missing. 
In this section, we present our DFD metamodel (\textbf{C1}). 
We also provide tooling centered around our metamodel, which aims to aid in the creation of new DFDs, the import of already existing DFD notations \cite{Schneider_Scandariato_2023, schneider2023microsecend}, and the automated derivation and visualization of DFDs from system architecture models like the PCM \cite{reussner2016a}.

\subsection{Unified Data Flow Diagram Metamodel}
\label{sec:dfd:model}
\begin{figure}[t]
    \centering
    \setlength{\abovecaptionskip}{3pt}
    \setlength{\belowcaptionskip}{-3pt}
    \includegraphics[width=\textwidth]{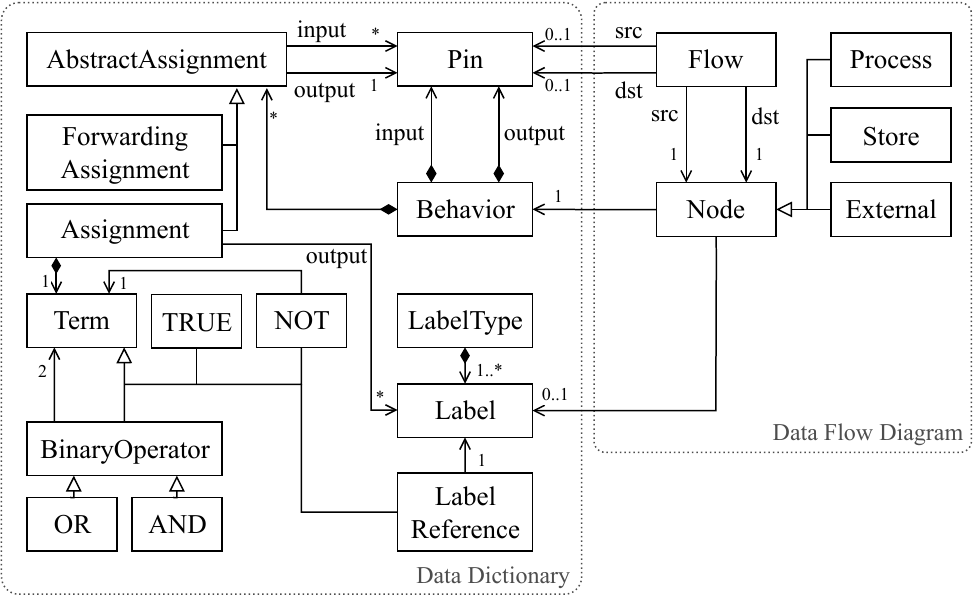}
    \caption{Metamodel of data flow diagrams and data dictionaries.}
    \label{fig:Metamodel}
\end{figure}

DFDs, as proposed by \textcite{demarco1979structure}, can be represented as DAGs showing the data flow and processing in software systems. 
\emph{Nodes} in these graphs represent \emph{External} entities like users, \emph{Processes} that can alter data, or \emph{Stores} like databases, connected by \emph{Flows} of data. 
\textcite{seifermann2022detecting} extend the notation by integrating several strands of work from different research groups into one unified metamodel. 
\autoref{fig:Metamodel} shows our metamodel that aligns with the unified DFD notation.
It is split into the so-called \emph{Data Dictionary} \cite{demarco1979structure} and the DFD.
The \emph{Data Dictionary} does not directly depend on a modeled system and can thus be reused while DFD elements are specific to a certain system.

The central part of the unified notation is the representation of behavior and characteristics as first-class entities. 
\emph{Labels} represent characteristics in the DFD, e.g., specifying the sensitivity of data, or the role of a user. 
They can either be defined as a characteristic of a \emph{Node} or as a characteristic of data flowing between \emph{Nodes}. 
\emph{Labels} are grouped in \emph{LabelTypes}.
The \emph{Behavior} of \emph{Nodes} defines which \emph{Labels} flow from one \emph{Node} to the next via the connecting \emph{Flow}.
It is made up of \emph{Pins} and \emph{Assignments}. 
Input \emph{Pins} represent required interfaces and output \emph{Pins} represent provided interfaces of nodes.
If a node has a certain \emph{Behavior}, it also has the corresponding input and output \emph{Pins}. 
A \emph{Flow} connects two \emph{Nodes} by connecting an output \emph{Pin} of the source \emph{Node} to an input \emph{Pin} of the destination \emph{Node}.
\emph{Assignments} define which labels flow out of a node. 
They reference input and output \emph{Pins} of their corresponding \emph{Behavior} and aggregate all \emph{Labels} of the data flowing in through the input \emph{Pin}. 
By evaluating a logical statement defined in the assignment, it is determined how the incoming \emph{Labels} are changed and passed on via the referenced output \emph{Pins}, e.g., the encryption of data can be represented by an \emph{Assignment} that adds an \emph{encrypted} label to the flowing data.

For assignments, we define two subclasses: \emph{Assignment} which contains a freely definable logical \emph{Term} that is evaluated to decide if a set of \emph{Labels} is applied to the output \emph{Pin}. The \emph{ForwardingAssginment} does not define a logical term but specifies that all \emph{Labels} that flow into the input pins are combined and directly forwarded to the output pin. 
The logical terms can be nested with binary operators \emph{AND} and \emph{OR} and negated with \emph{NOT} to express different statements. \emph{LabelReferences} are evaluated by checking if the referenced \emph{Label} flows into the node through one of the input pins of the \emph{Assignment}. 
In this case, the \emph{LabelReference} evaluates to true, otherwise to false. 
The \emph{Assignments} of a \emph{Behavior} are ordered. If a \emph{Behavior} contains multiple \emph{Assignments}, first all \emph{ForwardingAssignments} are evaluated and the \emph{Labels} for each output \emph{Pin} are saved. Other \emph{Assignments} add or remove labels for their specific output \emph{Pin}, depending on if their \emph{Term} evaluates to true or false.
Once all \emph{Assignments} are evaluated, the \emph{Labels} flow to the next \emph{Node}.

\subsection{Manually Defining Data Flow Diagrams}%
\label{sec:dfd:tool}
Manual ways to define DFDs that go beyond drawing on either paper or in software are limited. 
With our approach, we therefore offer a ready-to-use web editor to manually define DFDs and means to import DFDs from other notations.

\begin{figure}[t]
    \centering
    \setlength{\abovecaptionskip}{3pt}
    \setlength{\belowcaptionskip}{-3pt}
    \includegraphics[width=\textwidth]{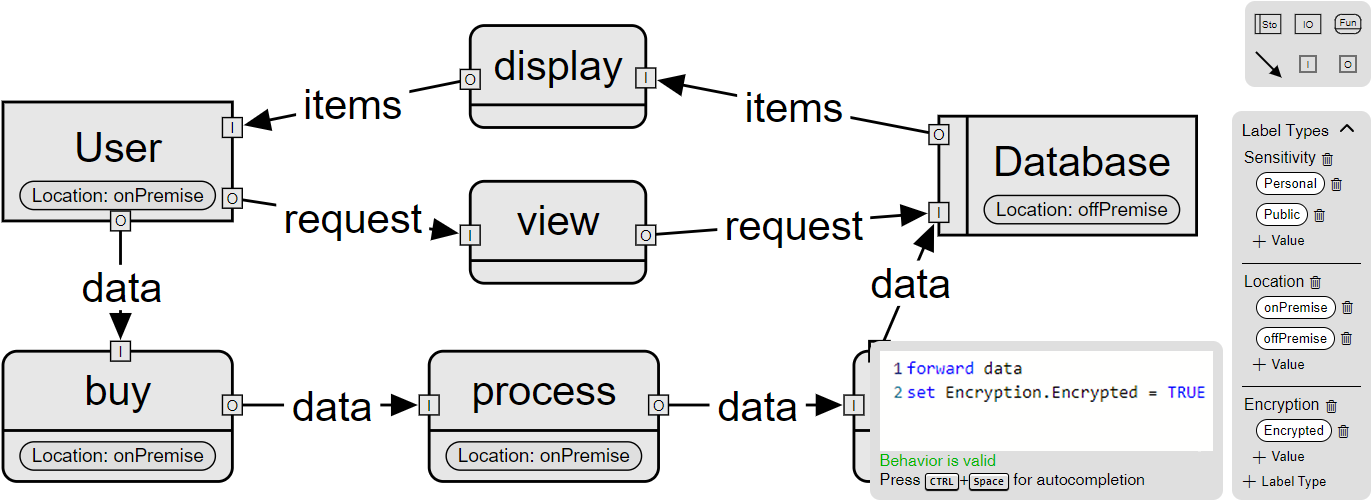}
    \caption{Screenshot of the web-based editor showing the DFD of a simplified online shop.}
    \label{fig:dfd:webeditor}
\end{figure}

Our web-based editor uses a notation that is compatible with the unified DFD notation from \autoref{fig:Metamodel}. 
We also incorporated the concept of the data dictionaries.
The graphical syntax follows earlier definitions of DFDs \cite{demarco1979structure,seifermann2022detecting}.
\autoref{fig:dfd:webeditor} shows the editor with an exemplary DFD. 
The toolbar on the right allows the creation of the three node types, data flow edges, and input and output pins via drag and drop. 
The \emph{Label Types} field allows the creation of label types and corresponding labels. 
Created labels can be annotated to a node by drag and drop.
Double-clicking on an output pin opens an editor for specifying assignments for the corresponding output pin.
Assignments can be defined in textual form using a DSL.
The \emph{forward} keyword is used to forward all labels of the corresponding input. 
The \emph{set} keyword is used to define an output label and a logical term, similar to the DFD metamodel.
Incoming data can be referenced via the name of the incoming edge. 
Labels are referenced by label type and label name.
Assignments are automatically syntax-checked, and issues are reported to the user. 
Additionally, our web editor supports highlighting in different colors and providing tooltips for nodes. This can be used to, e.g., visualize analysis results and provide additional information regarding identified security violations.

The manually created DFDs can be exported as JSON files.
To integrate the editor into our framework, we offer tooling that converts the JSON files of the web editor into an instance of our DFD metamodel.
The editor is implemented in TypeScript and uses the open-source diagramming framework Eclipse Sprotty. 
To ease the adoption of our approach, we additionally created extensible tooling for generating instances of our DFD metamodel from various inputs. 
At the time of writing, we support DFD notations in PlantUML and two different types of JSON notations.
To showcase this functionality, we have processed all security-enriched DFDs of the microSecEnD dataset of \textcite{schneider2023microsecend}. 
The resulting instances of our DFD metamodel can be found in our dataset \cite{dataset}.

\subsection{Automatically Deriving Data Flows from Architectural Models}
\label{sec:dfd:tool:derive}

Besides the manual modeling of DFDs, our framework also supports the automated extraction of data flows from the architecture description language PCM~\cite{reussner2016a}.
We choose PCM as it has already been used by previous data flow analysis approaches \cite{Seifermann2019, seifermann2022detecting}.
However, the described concept of data flow extraction is also applicable to other modeling languages like UML.

\begin{figure}[t]
    \centering
    \setlength{\abovecaptionskip}{3pt}
    \setlength{\belowcaptionskip}{-8pt}
    \includegraphics[width=0.9\textwidth]{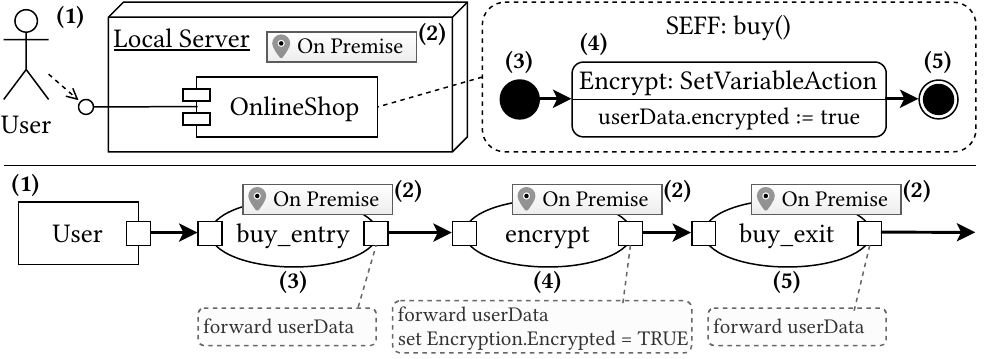}
    \caption{Simplified PCM model of the online shop example and the corresponding data flow with annotated node labels, data labels, and numbered transformation traces.}
    \label{fig:pcm2dfd}
\end{figure}

\autoref{fig:pcm2dfd} shows a simplified PCM model of an online shop.
It comprises information about components (e.g., \emph{Online Shop}), resources (e.g., \emph{Local Server}), deployment, usage, and system behavior as Service Effect Specification (SEFF) \cite{reussner2016a}.
The model is annotated with confidentiality-related labels that represent characteristics of data storage like \emph{On Premise} and data processing like the encryption of \emph{userData} in the \emph{SetVariableAction}.
In the lower half, we show the extracted data flow.
We annotate numbers to represent the transformation traces from PCM to DFD.
Note, that this is only a simplified example; realistic software systems contain more than one data flow and several hundred nodes \cite{hahnerSeams}.

Every action in the usage and system behavior is transformed into one DFD node.
This includes calls from the user, external calls between components, start and end nodes, and internal data processing nodes.
The nodes' pins correspond to the in and outgoing data types, e.g., \emph{userData}.
For every node, we perform a lookup of node labels, which can be annotated, e.g., to resources, or usage scenarios.
An exemplary lookup in the PCM model goes from the \emph{encrypt} node to the \emph{Online Shop} component via the deployment to the \emph{Local Server} resource which is annotated with \emph{On Premise}.
Additionally, we convert the modeled system behavior to assignments of our DFD metamodel like the encryption of \emph{userData} can be expressed.
The default case is the forwarding of labels.

The transformation considers all information that is relevant for security analysis, e.g., data processing and characteristics.
Other information is not transformed, e.g., components and servers do not cause additional elements in the DFD.
This enables a system view from the perspective of the data which is especially suitable for properties like confidentiality \cite{Seifermann2019}.
However, we store all traces to the originating PCM elements during the transformation.
This enables the evaluation of advanced queries and constraints in the data flow analysis.

\section{Data Flow Analysis}
\label{sec:analysis}

The original Prolog-based analysis of \textcite{seifermann2022detecting} realized the extraction of data flows and propagation of labels by first transforming the PCM models to an explicit DFD metamodel notation, then transforming the DFD elements to Prolog. Data flow constraints were checked by defining Prolog queries that are unique to the modeled system. As one DFD element with characteristic labels is transformed into multiple Prolog statements, the Prolog code grows exponentially with the model size. The exponential growth results in high demand for memory, as the whole Prolog program needs to be fully loaded by the Prolog interpreter.
As the analysis is made up of multiple chained transformations and intermediate model representations, the maintenance of the analysis was made even harder.

\noindent
Additionally, the approach of modeling data flows via logical statements in Prolog can lead to increased runtimes: Due to the lazy evaluation of Prolog, the Prolog-based analysis needs to reevaluate the characteristic labels of nodes for each different constraint.
For cases where very few nodes need to be evaluated, this might be an advantage. However, in using the analysis, the use case rarely occurs. For most constraints, like Role-based Access Control (RBAC), the node and data characteristic labels need to be evaluated at each node.


Due to the aforementioned reasons, we chose to implement the data flow analysis in Java and made the analysis more extensible as a central part of our framework. In this section, we first provide a general overview of the architecture of the analysis and provide a more detailed technical description of the extraction of data flows into flow graphs, label propagation, and constraint definition.

\subsection{Architecture Overview}
\label{sec:analysis:architecture}
\begin{figure}
    \centering
    \setlength{\abovecaptionskip}{3pt}
    \setlength{\belowcaptionskip}{-3pt}
    \includegraphics[width=.9\textwidth]{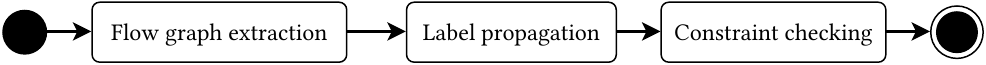}
    \caption{Analysis architecture as performed key activities.}
    \label{fig:ImplStructure}
\end{figure}

Our data flow analysis follows the general architecture of the Prolog-based data flow analysis of \textcite{seifermann2022detecting}. \autoref{fig:ImplStructure} shows the analysis steps and their sequential order as an activity diagram. 
Initially, the input models are loaded and references between model elements are resolved. 
This is done automatically by the Eclipse Modeling Framework (EMF).
Using the information from the models and annotations, we extract a set of \emph{transpose flow graphs (TFGs)} that each represents one unambiguous flow of data to a data sink in the modeled software architecture, i.e., the transpose rooted directed graph, where the root is a single data sink.
The extraction starts at each identified data sink and follows the modeled flow of data in the opposite direction.
Afterward, we transpose the graph to represent data flows between the vertices of the graph, so each TFG connects one or multiple data sources with a single data sink.
Each vertex represents one individual data processing step. 
If the analysis encounters an ambiguity in the data flow of the current element, it is resolved by creating copies of the current TFG, for each of the possible flows.
After all TFGs are extracted, we first evaluate the node characteristic labels of the vertices. Afterward, we propagate the data characteristic labels along the edges of the TFGs. 
Starting with the sink vertices, we calculate the data characteristics flowing into the current vertex by recursively evaluating the behaviors of the previous vertices in the TFG and tracing back the results.
How sinks are identified, how characteristic labels and vertex behavior are specified, and how they are evaluated, is specific to the input model type, e.g., DFD or PCM.

Using a set of fully propagated TFGs, data flow constraints can be checked. For example, by comparing propagated data characteristics with specified node characteristics, as described in \autoref{sec:dfd}.

\subsection{Flow Graph Extraction}

We specify extraction logic for creating TFGs and specific subclasses of \emph{vertex} for each element that represents a data flow node in DFDs and PCM models.

For the DFDs described in \autoref{sec:dfd:model}, sinks are nodes that either have no outgoing flows or nodes whose assignments for an output pin are independent of all its input pins.
Starting with these, the analysis performs a depth-first search over the DFD and creates vertices for each node. Ambiguities in the data flow exist if two or more flows point to the same input pin. To resolve the ambiguity, the analysis creates a copy of the current TFG for each path to the pin.

In the PCM, sinks are represented by the last element in usage scenarios. As the information regarding data flows is distributed across all PCM models, the analysis has to iterate over them and resolve relationships between elements. The analysis creates a vertex for all elements that can be annotated with a node characteristic label, that specifies data flow behavior, or that joins the control flow after a branch. The latter also creates a new TFG.
Calls to Service Effect Specifications (SEFFs) that are defined in interfaces are also handled separately: For each call of these SEFFs, a \emph{calling} and \emph{returning} vertex is created, which enclose the data flows, i.e., the vertices, which make up the SEFF internally.

\subsection{Label Propagation}
We individually propagate the characteristic labels for each TFG.
Each vertex references the input model element it represents and contains all logic regarding the calculation of node and data characteristic labels. First, we calculate the node characteristic labels and store them in the corresponding vertex. 
Starting from the sink of the TFG, we calculate the data characteristic labels that represent the output of the vertex and also store them in the corresponding vertex. 
This is achieved by recursively calling the calculation logic of all previous vertices and using the hereby calculated output labels as input.
Note, that we do not consider cycles in the propagation logic because TFGs represent DAGs.

For our DFD metamodel, the calculation of node characteristic labels is trivial, as DFD nodes already contain these labels, and vertices directly represent nodes. During the calculation of data characteristic labels, each vertex first recursively evaluates its input, as described above, iterating the DFD nodes that are connected by input pins. After the input has been evaluated the labels are aggregated and saved as output of the corresponding output pin.


In PCM, node characteristic labels are directly annotated to PCM elements like resource containers or usage scenarios. For the calculation of node characteristic labels, the vertex iterates over the relationships of the PCM element it represents and stores the annotated characteristic labels relevant to the vertex. For the calculation of data characteristic labels, the PCM-specific vertices use the output of the previous vertex in the TFG. In contrast to our DFDs, the PCM does not support the definition of multiple individual data flows between two nodes that each represents a separate flowing data variable. Rather, one flow between two vertices in the TFG encapsulates all data flowing between two nodes. To evaluate the input, the vertices filter the variables with their data characteristics to only include variables that are in the scope of the element represented by the vertex. To calculate the output data characteristic labels, the vertex evaluates stochastical expressions that are used in the PCM to define propagation behavior. 


\subsection{DSL-based Constraint Checking} 
\label{sec:analysis:dsl}
To help in the specification and checking of constraints and queries, we define a simple domain-specific language (DSL). 
We follow the general structure of the DSL by \textcite{Hahner2021b}, which was defined for the original analysis of \textcite{seifermann2022detecting} but simplify the approach by implementing it in Java and fitting it to our new implementation of the analysis.

\begin{figure}
    \centering
    \setlength{\abovecaptionskip}{3pt}
    \setlength{\belowcaptionskip}{-3pt}
    \includegraphics[width=0.89\textwidth]{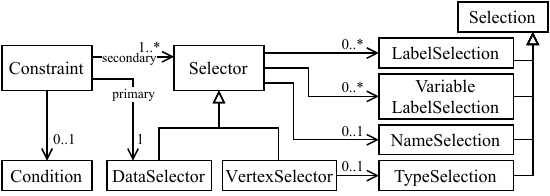}
    \caption{Metamodel showing the abstract syntax of the DSL for the data flow analysis.}
    \label{fig:dsl:class}
\end{figure}
\noindent
\autoref{fig:dsl:class} shows the abstract syntax of our DSL. 
A \emph{Constraint} is made up of primary and secondary selectors, as well as an optional condition. 
\emph{Selectors} are either specific for data or vertices. 
\emph{VertexSelectors} match the properties of the vertices themselves, while \emph{DataSelectors} match the propagated data characteristic labels of each vertex. 
They contain a set of \emph{Selections} that each represent a property. 
A \emph{Selection} can for example define a characteristic label or the name of a vertex or data. 
The \emph{VariableLabelSelection} does not reference a specific label but defines a variable that contains all labels of a given label type that are present at either the vertex or data. 
These variables can be compared in the \emph{Condition} of the constraint using set theory.
Executing the constraint searches all TFGs in the modeled software architecture using the flow graph extraction, propagates all labels, and tests each vertex. 
The selectors return all vertices in a TFG that match the properties defined by its selections. 
Constraints define a \emph{never flows} relationship between the primary \emph{DataSelector} and secondary selectors. 
The results of the primary and secondary selectors represent violations. 
If a condition is defined, it is evaluated in addition. 
In this case, the violations are the results of primary and secondary selectors, for which the condition evaluates to true.

 \begin{lstlisting}[
     float,
     belowcaptionskip=-5pt,
     mathescape=false,
     caption={Code snippet showing a DSL constraint for a simplified online shop.},
     label={lst:codesnippet}
 ]
   var analysis = new DataFlowAnalysisBuilder().build(); // simplified

   var constraint =  new Constraint()
                            .ofData()
                            .withLabel("Sensitivity", "Personal")
                            .withoutLabel("Encryption", "Encrypted")
                            .neverFlows()
                            .toVertex()
                            .withLabel("Location", "offPremise")
                            .create();
                     
   var violations = constraint.execute(analysis);
 \end{lstlisting}

\autoref{lst:codesnippet} demonstrates the concrete syntax of our DSL for the online shop example from \autoref{sec:dfd:tool}.
We provide a \emph{builder} to set up the analysis with required inputs, which is simplified in line 1.
We define a constraint using our DSL, starting in line 3.
For our example, we specify that \emph{personal data} (line 5) that is \emph{not encrypted} (line 6) should \emph{never flow} to vertices that are \emph{off-premise} of the online shop (line 9).
We execute the constraint in line 12. After the execution, the variable \emph{violations} contains a list of all constraint-violating vertices within the modeled software architecture.
If no violation has been found, the list is empty.

\section{Analysis Framework Extensions}
\label{sec:extension}

We demonstrate the extensibility of our framework with several related work \cite{boltz2023designing,boltz2022model,boltz2022handling,codeql,hahnerSeams,Hahner2023Uncertainty,cssda,walter2021architectural} that is either compatible to or already using our approach.

\textcite{boltz2023designing, boltz2022model} showcase the extension of both modeling and analysis for data protection and privacy.
As shown in \autoref{fig:overview}, they provide a GDPR metamodel and transformations from PCM and to and from our DFD metamodel.
Regarding the consideration of uncertainty within the software architectural design and system environment, multiple black-box and white-box extensions exist.
\textcite{walter2021architectural} use the data flow analysis as black-box together with PerOpteryx \cite{peropteryx} for design space exploration regarding confidentiality under structural uncertainty.
Other white-box extensions analyze access control under uncertainty \cite{boltz2022handling} or trace confidentiality violations to related uncertainty sources \cite{Hahner2023Uncertainty}.
Our framework is also used in an uncertainty impact analysis \cite{hahnerSeams} that predicts the impact of uncertainty on confidentiality based on the extracted data flows and a classification of uncertainty regarding confidentiality \cite{hahnerClassificationUncertainty}.





\section{Evaluation}
\label{sec:eval}
In our evaluation, we compare our new Java-based analysis to the Prolog-based analysis of \textcite{seifermann2022detecting}. The primary goals of this evaluation were to assess the accuracy and scalability of both analyses and to show that our Java-based analysis not only maintains the core functionalities of the Prolog-based analysis but also improves execution times and resource efficiency. Due to the lack of support for our new DFD metamodel in the Prolog-based analysis, our evaluation only focuses on PCM model instances. The evaluation of our analysis with a focus on the DFD metamodel or the extensions from \autoref{sec:extension} are considered potential future work.

\subsection{Evaluation Design}
To compare accuracy, we check whether both analyses correctly identify violations across various case study-based PCM models.
To ensure a good base for comparison, we utilize the same case study-based models employed by \textcite{seifermann2022detecting} for evaluating the accuracy of the Prolog-based approach.
The selected case studies use the default call return semantics of the current stable PCM version.
We executed both analyses with semantically equivalent constraint queries, using the count of accurately identified violations as the evaluation metric.

To examine and compare scalability, we measured the full execution time of both analyses while analyzing models of increasing size. To isolate the impact of distinct model features on scalability, we generated individual minimal models incrementally increasing the number of node characteristic labels, characteristic label propagations, variable actions, or SEFF parameters. We chose these elements, as they have the highest impact on either the length of Prolog code or Java loop iterations, depending on the analysis.
Each analysis was executed with a constraint designed to detect a violation at each node, thus ensuring a worst-case execution time scenario for both analyses.
For each run, we increase the model feature under consideration by the power of ten, starting at $10^0$ and ending with $10^5$.
We conducted each test 10 times and calculated the median execution time to mitigate outliers or measurement anomalies.
The analyses were performed on a dedicated VM equipped with 4 AMD Opteron 8435 cores, 97 GB RAM, running Debian 11 with OpenJDK 11/17.

\subsection{Evaluation Results}
In terms of accuracy, both analyses successfully identified the 42 violations present in the case study-based models without returning any false positives. \autoref{tab:Validation} shows the results of the accuracy evaluation and size of analyzed models. As both analyses performed the same, we assume, that our Java-based analysis is functionally equivalent to the Prolog-based analysis, when analyzing models using the call return semantics of the PCM.
\begin{table}
    \centering
    \setlength{\abovecaptionskip}{8pt}
    \setlength{\belowcaptionskip}{-5pt}
    \begin{tabularx}{\textwidth}{l>{\centering\arraybackslash}X >{\centering\arraybackslash}Xcc} 
        \toprule
        Case Study & Prolog-based & Java-based & Components & Labels \\
        \midrule
        ContactSMS \cite{Katkalov2017} & 10 violations & 10 violations & 3 & 4 \\
        FlightControl \cite{seifermann2022detecting} & 0 violations & 0 violations & 6 & 6\\
        FriendMap \cite{Tuma2019} & 0 violations & 0 violations & 5 & 12 \\
        Hospital \cite{Tuma2019} & 0 violations & 0 violations & 4 & 12 \\
        ImageSharing \cite{seifermann2022detecting} & 0 violations & 0 violations & 1 & 9 \\
        PrivateTaxi \cite{Katkalov2017}  & 0 violations & 0 violations & 13 & 20 \\
        TravelPlanner \cite{Katkalov2017} & 32 violations & 32 violations & 7 & 8 \\
        WebRTC \cite{Tuma2019} & 0 violations & 0 violations & 20 & 12 \\
        \bottomrule
    \end{tabularx}
    \caption{Accuracy results of both analyses compared and size of the models.}
    \label{tab:Validation}
\end{table}

\usepgfplotslibrary{groupplots}
\usetikzlibrary{matrix}
\begin{figure}[t]
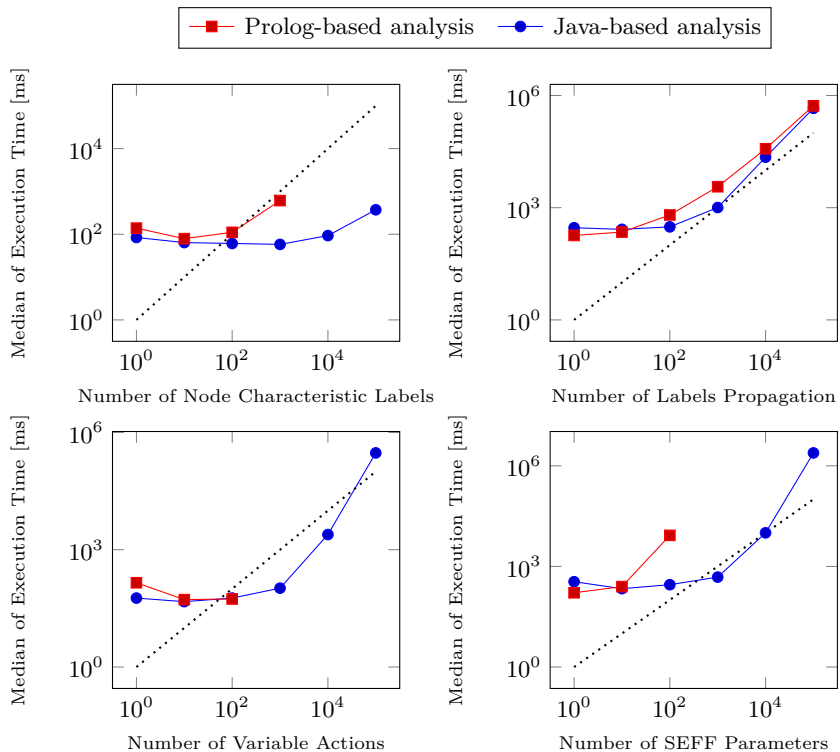

    \centering
    \setlength{\abovecaptionskip}{3pt}
    \setlength{\belowcaptionskip}{-3pt}
\begin{tikzpicture}
    \begin{groupplot}[group style={group name = graphs,group size = 2 by 4, horizontal sep=2cm, vertical sep=1.2cm}, height=5cm, width=5.4cm]
        \input{graphs/NewNodeCharacteristics}
        \input{graphs/NewCharacteristicsPropagation}
        \input{graphs/VariableActions}
        \input{graphs/SEFFParameter}
        \coordinate (bot) at (rel axis cs:1,0);
    \end{groupplot}
    \path (graphs c1r1.north west|-current bounding box.north)--
      coordinate(legendpos)
      (graphs c2r1.north east|-current bounding box.north);
    \matrix[
        matrix of nodes,
        anchor=south,
        draw,
        inner sep=0.2em,
        draw
    ]at([yshift=1ex]legendpos) 
    {
        \ref{plots:OldAnalysis}& Prolog-based analysis&[5pt]
        \ref{plots:NewAnalysis}& Java-based analysis\\};
\end{tikzpicture}
    \caption{Scalability results of the Prolog-based analysis and the Java-based analysis.}
    \label{fig:graph}
\end{figure}

\noindent
Regarding scalability, we plotted the results of both analyses as line graphs for each examined model feature, shown in \autoref{fig:graph}. Each graph contains data points from both analyses—the Prolog-based analysis (in red) and the Java-based analysis (in blue). Both axes are scaled logarithmically, with the x-axis showing the increasing number of model elements and the y-axis the median execution times in milliseconds.
Our evaluation showed that the Prolog-based analysis fails to complete a run for more than 1000 node characteristic labels or 100 for variable actions and SEFF parameters, due to high memory demand (see \autoref{sec:analysis}). 
In our tests, the analysis ran in \emph{out of memory errors} or crashed, despite the substantial \emph{97 GB} of available memory.
Regarding execution time behavior, while the Prolog-based analysis displayed an exponential increase in execution times or incomplete analysis runs for larger models, our Java-based analysis maintained nearly constant execution times up to $10^3$ elements for most evaluated cases. When increasing the number of label propagations, the execution time behavior of both analyses is similar.
The exponential increase in execution time of the Java-based analysis for larger models can be explained due to inefficiencies in TFG finding, and overhead during label propagation.

Overall, despite the noted increase in execution times for larger models in the Java-based analysis, we consider the time required in all scenarios feasible for design-time analyses. 
Our Java-based analysis, compared to the Prolog-based analysis, offers more manageable execution times and the capability to analyze large models, rendering it more suitable for real-world systems.

As parts of our evaluation are based around artificial scenarios and case studies, we discuss the external, internal, and construct validity, as well as reliability of our evaluation, as characterized by \textcite{runeson2012case}. Our main threat to external validity is the limited generalizability due to the case study-based evaluation. We try to mitigate this threat by using well-known case studies from literature to evaluate and compare accuracy. For the evaluation of scalability, the models were programmatically generated to only scale and focus on individual aspects of the models and analysis. 
A threat to the internal validity of our evaluation of scalability is that, due to the use of different technologies in both analyses, it was not possible to use the exact same constraints. We mitigate this threat by defining semantically equivalent constraints that find a violation at each node.
Our main threat to construct validity of our scalability evaluation is that it does not comprehensively cover all aspects that influence the execution time. We cannot fully mitigate this threat but have chosen the examined aspects based on the execution logic of both analyses and a previous scalability evaluation of \textcite{seifermann2022detecting}.
To mitigate threats regarding the reliability of our evaluation and to address the lack of replication packages in software architecture research \cite{Konersmann}, we have published a data set \cite{dataset}. The dataset contains all raw and compiled code artifacts, as well as an Eclipse-based product that already includes the plugins that make up the framework. The product can be used to model DFD or PCM instances and analyze them using our data flow analysis. We also include the raw results of our scalability evaluation and the used case study models.
\section{Conclusion and Future Work}
\label{sec:conclusion}
In this paper, we have presented our open and extensible framework for data flow analysis. We have introduced a unified DFD metamodel as a primary software architecture modeling artifact and input for our data flow analysis framework. We have described means that we provide to manually define DFDs as well as automatically derive them from the architecture description language PCM and other third-party representations.

Based on the approach of \textcite{seifermann2022detecting}, we have implemented a Java-based data flow analysis. We described the general architecture of the analysis and provided detailed technical descriptions of the core features. For the analysis, we have defined an extensible intermediate representation of data flows, called transpose flow graphs. We have described how data flows are extracted from input models and how characteristic labels are propagated using our new intermediate representation. To enable the definition of data flow constraints for the analysis, we have defined a new domain-specific language. 

We highlight the problems of the Prolog-based analysis of \textcite{seifermann2022detecting} and show in our evaluation, that our Java-based analysis is functionally equivalent to the Prolog-based analysis and can analyze larger system models.

\noindent
In future work, we aim to further enhance the tooling that makes up our framework.
We also aim to further work on the various framework extensions, like the data protection \cite{boltz2022model} and uncertainty analyses \cite{Hahner2023Uncertainty} 
and include more cooperation points of our framework, e.g., with continuous security analysis \cite{cssda}. Lastly, we aim to comprehensively evaluate the overall approach of our framework.

\subsection*{Acknowledgements}
\small{This publication is partially based on the research project SofDCar (19S21002), which is funded by the German Federal Ministry for Economic Affairs and Climate Action. This work was also supported by funding from the topic Engineering Secure Systems of the Helmholtz Association (HGF) and by KASTEL Security Research Labs, the BMBF (German Federal Ministry of Education and Research) grant number 16KISA086 (ANYMOS), and the NextGenerationEU project by the European Union (EU).
We like to thank Felix Schwickerath, Tom Hüller, Daniel Huber, Tizian Bitschi, Anne-Kathrin Hermann, and Nils Niehues for their support in the development of the presented work.}

\printbibliography[heading=bibintoc]

\end{document}